\documentclass[12pt,a4paper]{article}
\usepackage{amsmath}
\usepackage{amsfonts}
\usepackage{amssymb}
\usepackage{textcomp}
\usepackage[dvips]{graphicx}
\usepackage{epsfig}
\usepackage{bm}
\usepackage{dcolumn}
\usepackage{amsfonts}
\usepackage{color}
\usepackage{lineno}
\usepackage{empheq,etoolbox,float}
\usepackage{subcaption}

\usepackage[left=2cm,right=2cm,top=2cm,bottom=2cm]{geometry}

\begin{document}
\title{\textbf{Slow-roll approximations for Gauss-Bonnet inflation revisited}}
\author{Bogdan A.~Rudenko$^{a,b}$\footnote{E-mail: rudenko.sci@gmail.com},  Maria A.~Skugoreva$^{a}$ \footnote{E-mail: masha-sk@mail.ru}, and  Alexey V.~Toporensky $^{a}$ \footnote{E-mail: atopor@rambler.ru}
\\
\small$^a$ Sternberg Astronomical Institute, Lomonosov Moscow State University, \\ \small Universitetsky Prospect~13, Moscow 119991, Russia\\
\small$^b$ Faculty of Physics, Lomonosov Moscow State University,\\ 
\small Leninskie Gory 1, 119991, Moscow, Russia
}

\date{ \ }
\maketitle
\begin{abstract}
~~~~In our paper we consider the validity of slow-roll approximations for Gauss-Bonnet inflation introduced in~\cite{We}. In contrast to the cited paper where the coupling function before the Gauss-Bonnet term have been chosen as a decaying function of the scalar field, here we consider growing coupling functions. We have found that while in~\cite{We} new slow-roll approximations work considerably better, now they do not increase the precision. Moreover, we identify some cases where more involved approximations work worse than the standard one. Corresponding explanations of such a situation are given.
\end{abstract}
\maketitle

\section{Introduction}
~~~~The problem of finding adequate slow-roll approximations~\cite{We,Liddle1994,Akin:2020qjw} attracts considerable attention recently as the search for inflation models~\cite{Starobinsky1979,Guth1981,Linde1982,Linde1983,Mukhanov1981} reproducing observable characteristics of the CMB~\cite{PlanckCollaboration2023,Durrer2015,Abazajian2015,Louis2025,Calabrese2025,DESICollaboration2024} has been directed mostly towards modified gravity models~\cite{Capozziello2011,Lombriser2011}. Indeed, current observational bounds~\cite{Akrami2019,BICEP:2021xfz,Galloni:2022qjw} have already ruled out any power-law scalar field potentials, and the remaining alternatives are either very shallow potentials or involve going beyond General Relativity (GR). The latter choice requires reconsideration of slow-roll approximations~\cite{Liddle1994,Guo:2010jr}.

    In principle, all equations of motion can be solved numerically; however, the existence of analytical formulas usually enables qualitative analysis and makes the situation more transparent. On the other hand, we need not oversimplify the cosmological dynamics. Already in the case of a nonminimally coupled scalar field (with no other modifications of GR) two different approximations have been used in the literature, and one of them is considerably less reliable though being simpler one~\cite{Jrv2022}.

    Further deviations from GR, for example, considering a scalar field coupling not only with curvature, but also with the Gauss-Bonnet term~\cite{Chern1944,Jiang2013,VanDeBruck:2016jll,Pozdeeva2021,Pozdeeva:2019akw,Koh:2016abw,Odintsov:2023qjw,Gasperini_1997,Neupane_2006,Tsujikawa_2002,Antoniadis:1993jc,Cartier:2001is,Kawai:1998ab,Kawai:1999pw,Torii:1996xd,Hwang:2004th,Sami:2005zc,Tsujikawa:2006tv,Bahamonde:2017ize,Zhu2018,Zhu2023,Zhu2025,Odintsov2020,Odintsov2025,Oikonomou2021a,Oikonomou2021b,Oikonomou2022a,Oikonomou2022b,Oikonomou2024,Oikonomou2025,Yogesh2025a,Yogesh2025b,Yogesh2025c} makes the question of appropriate approximations even less clear. A direct analog of the approximation working better for curvature coupling shows very bad behavior for particular cases of Gauss-Bonnet coupling.

    New, more involved, approximations could cure this situation~\cite{We,We1}. However, going to more and more complicated approximations results in using less and less transparent formulas, so that advantages of analytical approach begin to be washed out.

    Motivated by this we consider in the present paper couplings with the Gauss-Bonnet term which are qualitatively different from those considered in~\cite{Pozdeeva:2020zpu}. In the cited paper the form of the coupling was motivated to reproduce observational parameters of the CMB with usual quadratic or quartic potentials (which is now impossible for a simple scalar field without any modifications), and have been chosen to be power-law with a negative power index. The goal of the present paper is to consider other forms of coupling functions and to check the validity of several types of slow-roll approximations introduced in recent papers~\cite{We,We1} rather  than to fit actually the experimental data. That is why some of our examples below will show the values of observational parameters out of present bounds. 

    The paper is organized as follows. In Section~2, we present the full theoretical framework in which our analysis is carried out. Section~3 presents the main results obtained from the numerical analysis, and Section~4 contains our conclusions.

\section{Theoretical framework}
\subsection{Inflation with nonminimal coupling}
~~~~We consider inflationary models with the Gauss-Bonnet term coupled to a scalar field, described by the action:

\begin{equation}
S = \int d^4x\sqrt{-g} \left( U R - \frac{1}{2} g^{\mu\nu} \partial_\mu \phi \partial_\nu \phi - V(\phi) + \xi(\phi) \mathcal{G} \right),
\label{eq:action}
\end{equation}
where $U=\frac{1}{16\pi G}=\frac{{(M_{\text{Pl}})}^2}{2}$ (further we use Planck units $c=\hbar=8\pi G=1$), $R$ is the Ricci scalar, $V(\phi)$ is the potential of the inflaton, $\xi(\phi)$ is the coupling function, and $\mathcal{G}$ is the Gauss-Bonnet invariant.

    The Gauss-Bonnet invariant has the form:
\begin{equation}
\mathcal{G} = R_{\mu\nu\rho\sigma}R^{\mu\nu\rho\sigma} - 4R_{\mu\nu}R^{\mu\nu} + R^2.
\label{eq:gb_invariant}
\end{equation}
In the spatially flat Friedmann-Lema\^{i}tre-Robertson-Walker (FLRW) metric with
\begin{equation}
ds^2 = -dt^2 + a^2(t) d\mathbf{x}^2,
\label{eq:flrw}
\end{equation}
where $a(t)$ is the scale factor, the following system of the equations of motion is obtained \cite{VanDeBruck:2016jll, Pozdeeva:2020zpu}
\begin{subequations}
\begin{empheq}[left=\empheqlbrace]{align}
& 12 H^2(U - 2\xi_{,\phi} \dot\phi H) = \dot{\phi}^2 + 2V, 
\label{eq4} \\
& -4 \dot H(U-2\xi_{,\phi}\dot\phi H) - {\dot\phi}^2 + 4\xi_{,\phi\phi} {\dot\phi}^2 H^2 + 4 \ddot\phi\xi_{,\phi}H^2 - 4\xi_{,\phi} \dot\phi H^3 = 0, 
\label{eq5} \\
& \ddot\phi+ 3H \dot\phi+ V_{,\phi} + 12\xi_{,\phi} H^2 \dot{H} + 12\xi_{,\phi}H^4 = 0. 
\label{eq6}
\end{empheq}
\end{subequations}
where $H\equiv\frac{\dot a}{a}$ is the Hubble parameter, dots denote the derivatives with respect to the cosmic time $t$ and $A_{,\phi}\equiv\frac{dA}{d\phi}$ for any function $A(\phi)$.

    Solving the system of the equations of motion (\ref{eq4})-(\ref{eq6}) we find the system for the numerical integration
\begin{subequations}
\begin{empheq}[left=\empheqlbrace]{align}
&\dot{\phi} = \psi, \\
&\dot{\psi} = \frac{1}{U-2\xi_{,\phi}\psi H+12(\xi_{,\phi})^2H^4}\Biggl(9\xi_{,\phi}\psi^2H^2-12\xi_{,\phi}\xi_{,\phi\phi}\psi^2H^4+36(\xi_{,\phi})^2\psi H^5+   \nonumber \\&\hspace{1.05cm}+2\xi_{,\phi}V_{,\phi}\psi H-U(3H\psi+V_{,\phi}+12\xi_{,\phi}H^4) \Biggr), \\
&\dot{H} = \frac{-\frac{\psi^2}{4}+\xi_{,\phi\phi}\psi^2H^2-4\xi_{,\phi}\psi H^3-\xi_{,\phi}V_{,\phi}H^2-12(\xi_{,\phi})^2H^6}{U-2\xi_{,\phi}\psi H+12(\xi_{,\phi})^2H^4},\\
&\dot{a} = aH, 
\end{empheq}
\label{eq13}
\end{subequations}
where three independent variables are $\phi$, $\psi$ and $a$. The equation $12U H^2={\psi}^2 +2V+24\xi_{,\phi}\psi H^3$ is checked at each step of numerical integration.

    Taking into account that $\frac{d}{dt} = H\frac{d}{dN}$, where $N=\ln (a/a_e)$ is the e-foldings number with $a_e=const$, we can rewrite the system (\ref{eq13})
\begin{subequations}
\begin{empheq}[left=\empheqlbrace]{align}
&\frac{d\phi}{dN} = \frac{\psi}{H}, \\
&\frac{d\psi}{dN} = \frac{\dot{\psi}}{H} = \frac{1}{H(U-2\xi_{,\phi}\psi H+12(\xi_{,\phi})^2 H^4)}\Biggl(9\xi_{,\phi}\psi^2 H^2-12\xi_{,\phi}\xi_{,\phi\phi}\psi^2 H^4+  \nonumber
\\&\hspace{2.2cm}+36(\xi_{,\phi})^2\psi H^5+2\xi_{,\phi}V_{,\phi}\psi H-U(3H\psi+V_{,\phi}+12\xi_{,\phi}H^4)\Biggr), \\
&\frac{dH}{dN} = \frac{\dot{H}}{H} = \frac{-\frac{\psi^2}{4}+\xi_{,\phi\phi}\psi^2 H^2-4\xi_{,\phi}\psi H^3-\xi_{,\phi}V_{,\phi}H^2-12(\xi_{,\phi})^2 H^6}{H(U-2\xi_{,\phi}\psi H+12(\xi_{,\phi})^2H^4)}. 
\end{empheq}
\label{eq:numerical-1}\
\end{subequations}
Here two independent variables are $\phi$, $\psi$ and the relation $12U H^2={\psi}^2 +2V+24\xi_{,\phi}\psi H^3$ is checked at each step of numerical integration.

  Therefore, we can use for the numerical integration any of the systems~(\ref{eq13}) or~(\ref{eq:numerical-1}).

\subsection{Slow-roll approximations}
~~~~As already mentioned in the Introduction, the primary approximation used in inflationary cosmology is the slow-roll approximation~\cite{Liddle1994}. It is based on the assumption that the inflaton field evolves very slowly compared to the characteristic timescale of changes in the system's parameters.

    Following Refs.~\cite{Guo:2010jr,Koh:2016abw,Odintsov:2023qjw,Pozdeeva:2020zpu}, we consider the slow-roll parameters:

\begin{equation}
\varepsilon_1 = - \frac{\dot{H}}{H^2} = - \frac{1}{2}\frac{d \ln\left(H^2\right)}{dN},\;\;\;\;\; \varepsilon_{i+1} = \frac{d \ln\left(|\varepsilon_1|\right)}{dN}.
\label{eq:epsilon}
\end{equation}
\begin{equation}
\delta_1 = - \frac{2}{U}H\dot{\xi} = - \frac{2}{U}H^2\xi_{,\phi}\frac{d \phi}{dN},\;\;\;\;\; \delta_{i+1} = \frac{d \ln\left(|\delta_1|\right)}{dN}.
\label{eq:delta}
\end{equation}

    The standard slow-roll conditions, imposed on the slow-roll parameters during inflation, retain their usual form:

\begin{equation}
|\varepsilon_i| \ll 1, \;~~~~ |\delta_i| \ll 1.
\end{equation}

    The main observable parameters, such as the amplitude of the scalar perturbation $A_s$ and the tensor-to-scalar ratio $r$, are related to the slow-roll parameters as follows~\cite{Guo:2010jr,We}:

\begin{equation}
r = 8|2\varepsilon_1 - \delta_1| = \left| \frac{U\delta_1^2}{\xi_{,\phi}^2 H^4}  - 8\delta_1\delta_2 + 8\delta_1\varepsilon_1 \right|,
\label{eq:r}
\end{equation}
\begin{equation}
A_s = \frac{H^2}{\pi^2 U r}.
\label{eq:a_s}
\end{equation}
The spectral index is given in the form:
\begin{equation}
n_s = 1 - 2\varepsilon_1 - \frac{2\varepsilon_1\varepsilon_2 - \delta_1\delta_2}{2\varepsilon_1 - \delta_1}.
\label{eq:n_s}
\end{equation}

\subsubsection{Standard slow-roll approximation}
~~~~In the context of inflation with the Gauss-Bonnet coupling, a set of approximate equations has been developed in Ref.~\cite{Guo:2010jr}. The leading order equations have the following form \cite{We}:
\begin{subequations}
\begin{empheq}[left=\empheqlbrace]{align}
H^2 &\approx \frac{V}{6U}, \\
\dot{H} &\approx -\frac{\psi^2}{4U} - \frac{\xi_{,\phi} H^3 \psi}{U}, \label{2.22b}\\
\psi &\approx -\frac{V_{,\phi} + 12\xi_{,\phi} H^4}{3H}.
\end{empheq}
\label{eq:classical-approx}
\end{subequations}

    The inflationary parameters $n_s(\phi) $, $r(\phi)$ and $A_s(\phi)$ can be computed using formulae~(\ref{eq:r})-(\ref{eq:n_s}).

\subsubsection{I new approximation}
~~~~Let us assume:
\begin{equation}
\delta_1 \ll 1.
\label{eq:small_parameter}
\end{equation}
If we expand the expression for \( H^2 \) in a series with respect to \( \delta_1 \) and retain terms up to second-order, another slow-roll approximation, retaining one more term in this expansion compared to the standard one can be constructed~\cite{We}:
\begin{subequations}
\begin{align}
H^2 &\approx \frac{V}{6U} + \frac{V}{6U} \delta_1 + \mathcal{O}(\delta_1^2) 
\label{eq:h2_expansion_first_order} \\
&\approx \frac{V}{6U} (1 + \delta_1).
\label{eq:h2_expansion_rewritten}
\end{align}
\label{eq:first-approx}
\end{subequations}

    Substituting the values for $\delta_1$ and replacing equation~(\ref{2.22b}) with the approximation the original equation~(\ref{eq5}), we obtain:

\begin{subequations}
    \begin{empheq}[left=\empheqlbrace]{align}
        H^2 &\approx \frac{1}{6U^2} \left( UV + 2V \xi_{,\phi} H \psi \right), \\
        \dot{H} &\approx -\frac{\psi^2 + 4H^3\xi_{,\phi}\psi}{4U(1-\delta_1)}. 
    \end{empheq}
    \label{eq:first-approx-system}
\end{subequations}
This approach is particularly useful for scenarios where a nonminimal coupling function $\xi(\phi)$ influences significantly the dynamics of inflation. By keeping an additional term in the expansion compared to the standard approximation, we account for corrections due to the Gauss-Bonnet coupling. 

\subsubsection{II new approximation}
~~~~The second more involved  approximation arises from a more accurate consideration of the effect of the Gauss-Bonnet term on inflation dynamics. Let us express \( H^2 \) from the dynamical system equation~(\ref{eq4}), rewriting it as:
\begin{equation}
12UH^2(1 - \delta_1) = \psi^2 + 2V.
\label{12uh}
\end{equation}
Multiplying \eqref{12uh} by \( H^2 \) and expressing \( \psi \) in terms of \( \delta_1 \), we obtain:
\begin{equation}
12U(1 - \delta_1)H^4 - 2VH^2 - \frac{\delta_1^2 U^2}{4\xi_{,\phi}^2} = 0.
\end{equation}
Neglecting the \( \delta_1^2 \) term, we get \cite{We}:
\begin{equation}
H^2 \approx \frac{V}{6U(1 - \delta_1)}.
\label{eq:second-approx}
\end{equation}
If we expand this expression in powers of $\delta_1$ and retain only the first two terms, we recover Approximation~I. This suggests that (\ref{eq:second-approx}) should generally provide better accuracy, as it includes higher-order contributions. However, Approximation~II leads to divergence in the Hubble parameter as $\delta_1 \to 1$, while Approximation~I remains finite. Therefore, in regimes where $\delta_1 \to 1$, Approximation~I may perform better despite being a lower-order expansion, due to its regular behaviour. This trade-off between accuracy and stability near $\delta_1 = 1$ motivates a comparative analysis of both approximations against numerical solutions.

\section{Numerical and approximated solutions}
~~~~In this section we perform numerical investigation of the exact system (\ref{eq:numerical-1}) and compare results with the approximations described in the previous section. We choose the initial data so that the number of e-foldings till the end of inflation exceeds observationally preferred value near $60$ by about $10$. This means that 
when the system passes $60$ e-foldings till the end of inflation, it is already at the inflationary 
trajectory and our actual details of the initial conditions are already washed out.

    We start our considerations with the massive scalar field potential 
\begin{equation}
\label{V}
V(\phi)=V_0\phi^2,
\end{equation}
choosing
\begin{equation}
\label{paramV}
V_0=1.5\cdot 10^{-11}.
\end{equation}
 Such a potential leads to $\dot\phi$ at the inflation stage quite large so that the Gauss-Bonnet corrections are non-negligible (we remind a reader that constant in time coupling function before the Gauss-Bonnet term makes it, as a topological invariant in $4$ dimensions, irrelevant for a cosmological dynamics). So that, our goal here is to study slow-roll approximations, but not to reproduce observational bounds on the tensor-to-scalar ratio $r$.\footnote{Note, that there are some approaches to evolutions of primordial inhomogeneities which remove the current bound on $r$, see, for example, \cite{Sudarsky}. That is why, apart from academic purposes, keeping in mind models disfavored only by $r$ might be useful.}
If we consider a coupling function growing with $\phi$ it is quite natural to choose a power-law behavior. Since the coupling with the Gauss-Bonnet term should be dimensionless, it is reasonable to express the coupling function as
\begin{equation}
\label{xi}
\xi(\phi)={(\phi/\phi_0)}^n.
\end{equation}
The parameter $\phi_0$ have the obvious meaning of a scale starting from which the coupling becomes important. When it is large enough, the Gauss-Bonnet term is not important during most part of inflation phase, and we can expect that the simpler slow-roll approximation  is enough to reproduce the inflationary dynamics. Indeed, it appears that for $\phi_0=1$ the relative differences between parameters calculated using different approximations are very small, being, for example, for $\varepsilon_1$ and $n_s$
  of the order of $10^{-10}$ for $n=2$  and of the order of $10^{-7}$ for $n=4$.

\subsection{Cases when the standard approximation works well}
\begin{figure}
\includegraphics[scale=0.28]{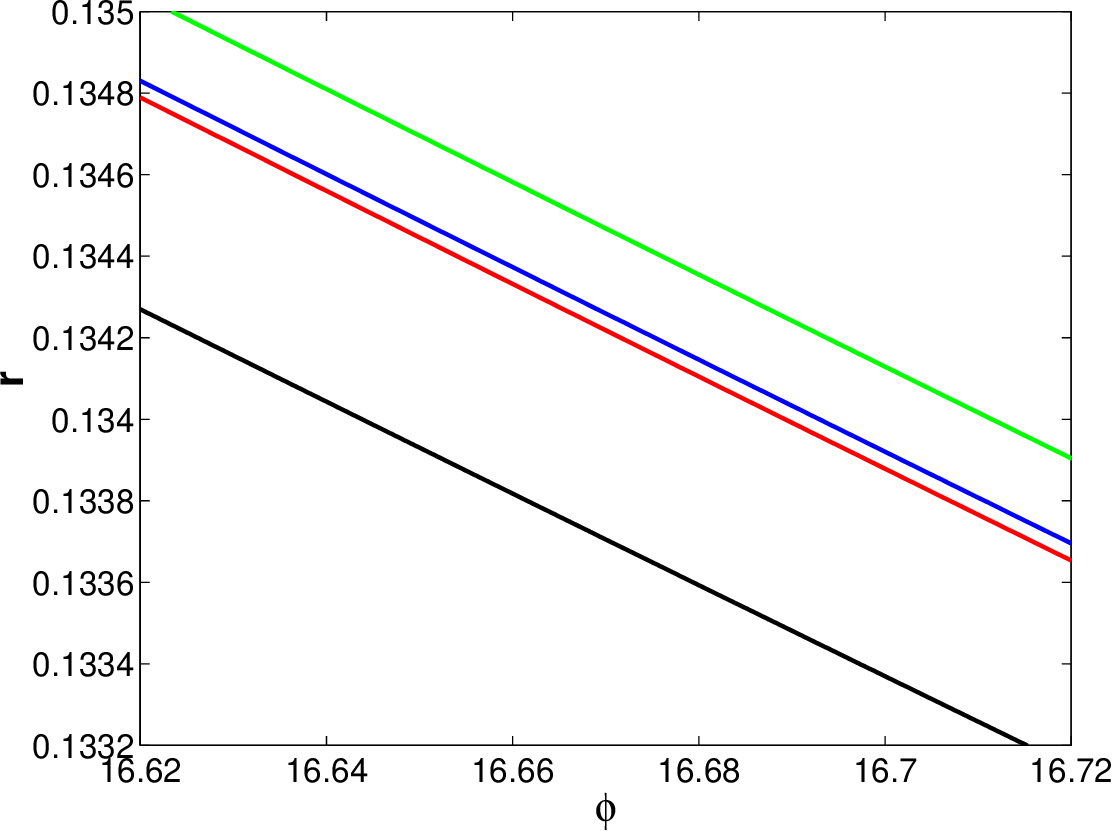}~~~~
\includegraphics[scale=0.28]{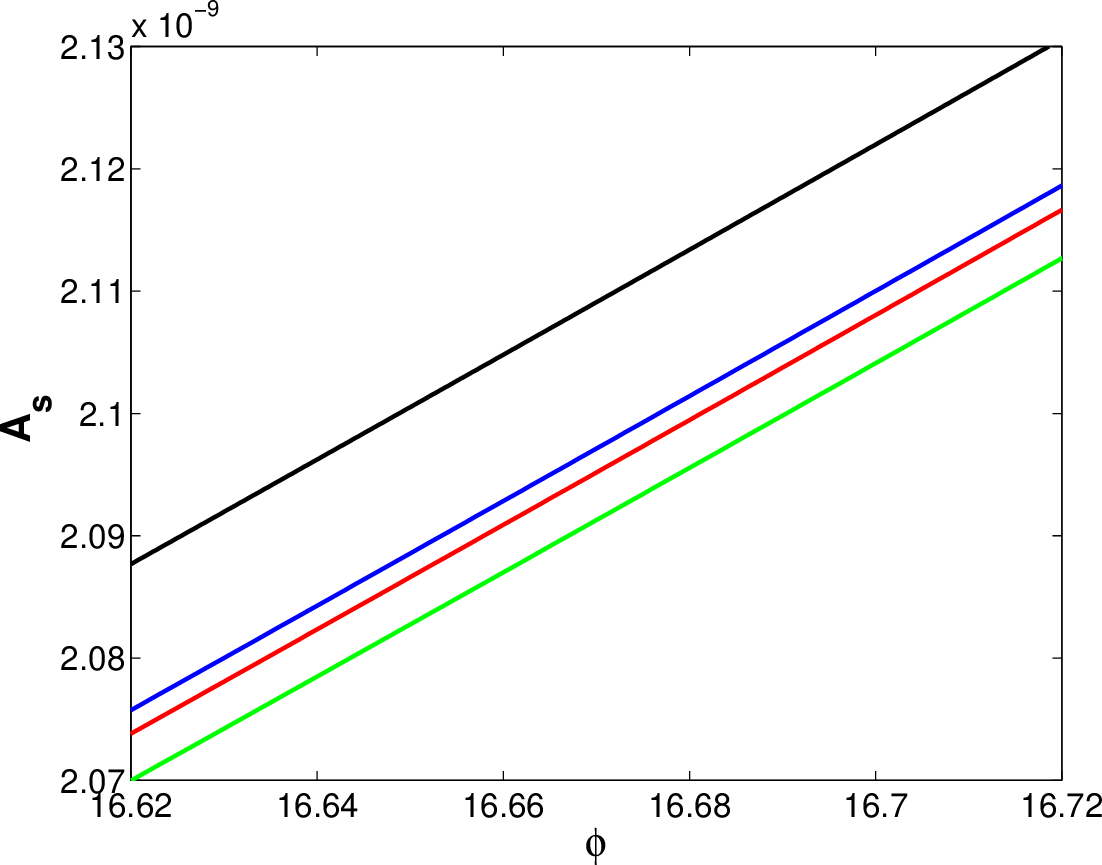}~~~~
\includegraphics[scale=0.28]{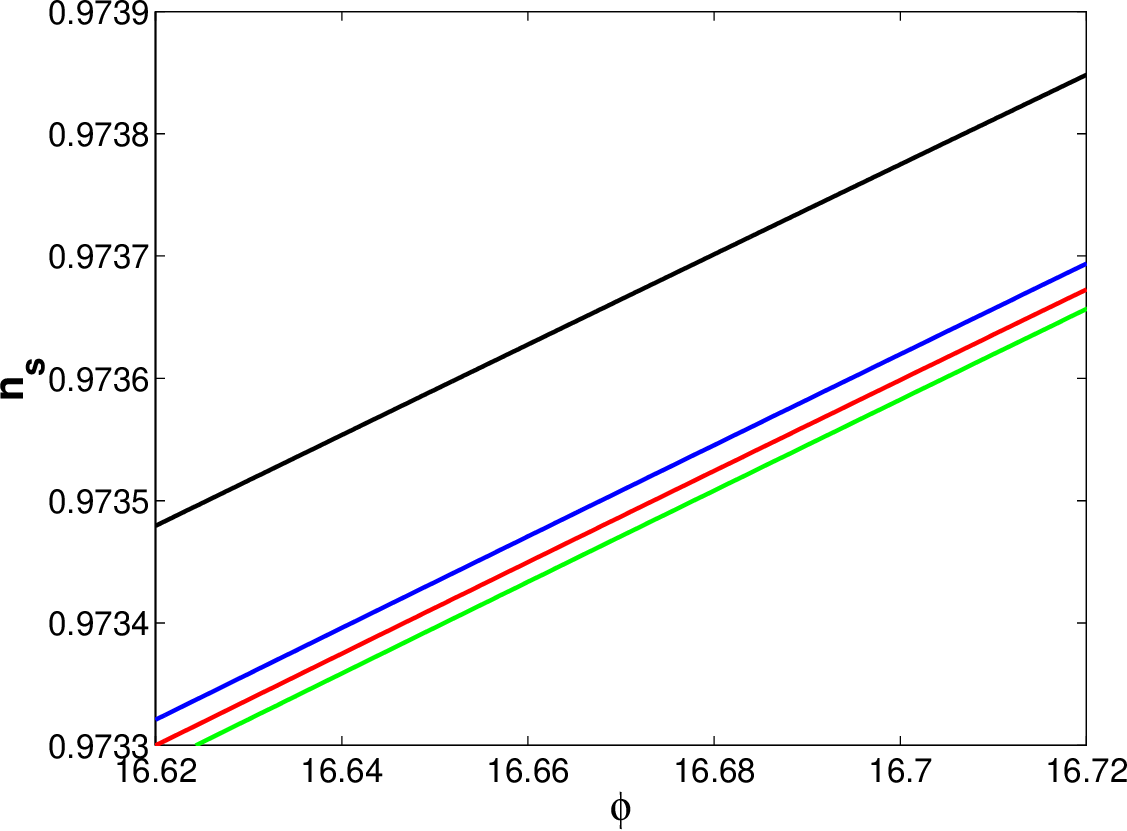}
    \caption{The plot depicts the evolution of the inflationary parameters $r$~(left), $A_s$~(middle), $n_s$~(right) at the onset of inflation for the model (\ref{V}), (\ref{paramV}), (\ref{xi}), (\ref{paramxi}). The black line represents the numerical solution of system~\eqref{eq:numerical-1}, blue --- classical slow-roll approximation, red --- new Approximation~I, green --- new Approximation~II.}
    \label{Fig1}
\end{figure}

~~~~With $\phi_0$ decreases, the difference between approximations grows. A typical example is  presented in Fig.~\ref{Fig1} where we use 
\begin{equation}
\label{paramxi}
\phi_0=4.4\cdot 10^{-3},~~~~n=2
\end{equation}
 for function (\ref{xi}). As we can see, old approximation does its job with a comparable accuracy as new approximations. This means that for such coupling functions in the regime when the Gauss-Bonnet corrections are small enough it is not necessary to use more complicated approximations. 

    What was unexpected to us, sometimes old approximation appears to be significantly better that the new ones. We detected this for asymptotically flat coupling function, and present below two examples of such a situation. Since an asymptotically flat function tends to a constant for large $|\phi|$, the Gauss-Bonnet term does not affect the high-energy dynamics and is important only when $|\phi|$ is small enough. An example of such function is 
\begin{equation}
\label{xi1}
\xi(\phi)=\frac{\phi^2}{(\alpha\phi)^2+\phi_0^2}
\end{equation}
where a constant $\alpha$ is chosen so as two terms in the denominator being of the same order for intermediate $\phi$. It is chosen to be $\alpha=3\cdot 10^{-6}$ for the results presented below.

\subsubsection{Example 1}
~~~~We consider a scenario with the potential (\ref{V}) choosing
\begin{equation}
\label{ex1paramV}
V_0=1.2\cdot {10}^{-11}
\end{equation}
and
the coupling function (\ref{xi1}) with 
\begin{equation}
\label{ex1paramxi}
\phi_0 = {10}^{-6},~~~~\alpha=3\cdot  {10}^{-6}.
\end{equation} 
To study the evolution of the inflationary scenario, we present a set of plots showing the behavior of the slow-roll parameters \(\varepsilon_1\), \(\delta_1\) and the inflaton field \(\varphi\) at the beginning and end of inflation, along with the key inflationary observables: the tensor-to-scalar ratio \(r\), the scalar amplitude \(A_s\), the scalar spectral index \(n_s\).
    
\begin{figure}
\includegraphics[scale=0.46]{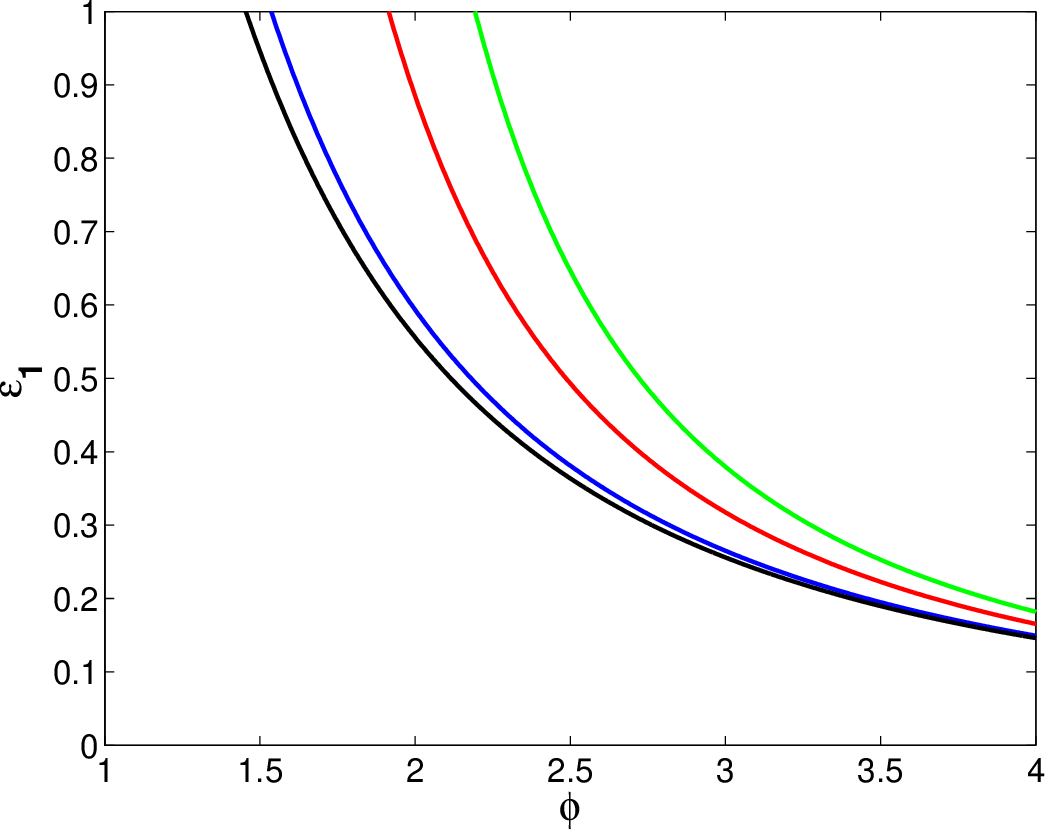}~~~~
\includegraphics[scale=0.46]{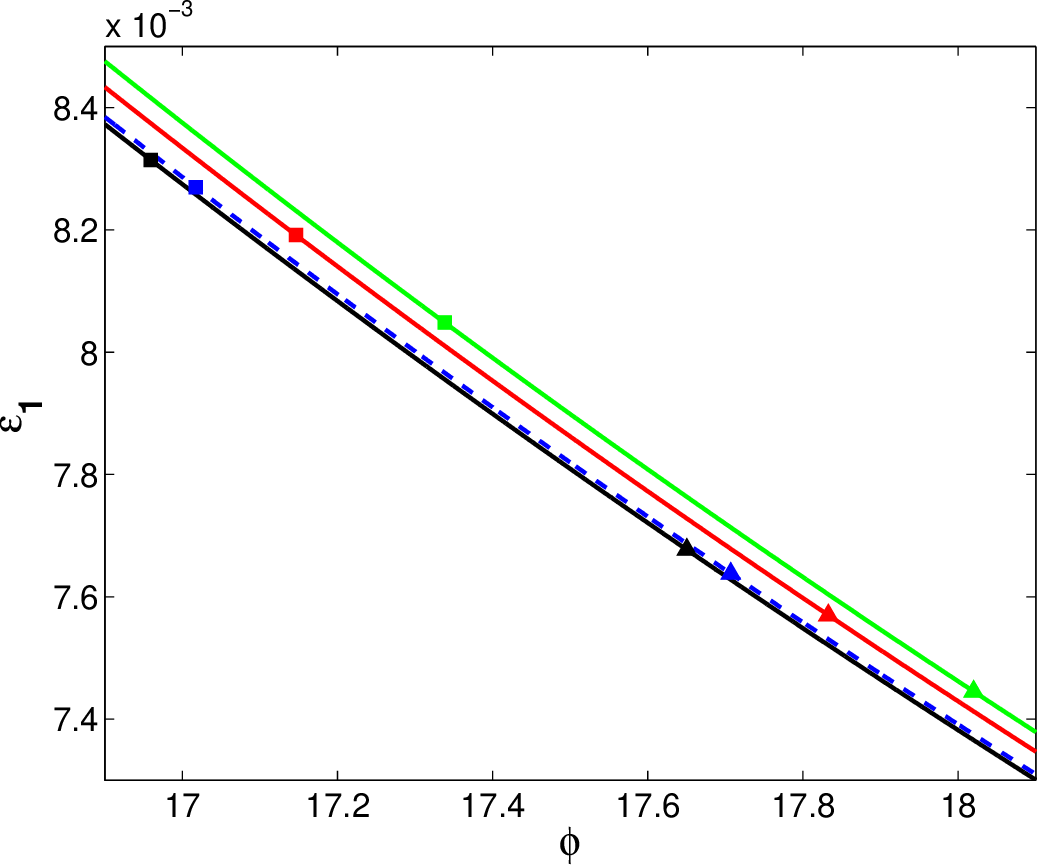}
\caption{\textit{On left panel}: The evolution of the slow-roll parameter $\varepsilon_1(\phi)$ at the end of inflation for the Example~1. The black line represents the numerical solution of system~(\ref{eq:numerical-1}), blue --- classical slow-roll approximation, red --- new Approximation~I, green --- new Approximation~I. For better visibility of the curves, the $\phi$-axis display range was reduced from $[0; 20]$ to $[1; 4]$. \textit{On right panel}: The evolution of the slow-roll parameter $\varepsilon_1(\phi)$ at the beginning of inflation for the Example~1. For comparative purposes, the plot indicates different numbers of e-foldings: \(N = 65\) (triangles) and \(N = 60\) (squares). For visual clarity, the blue curve representing the standard slow-roll approximation is shown as a dashed curve.}
\label{Fig2}
\end{figure}

    As seen in Fig.~\ref{Fig2}, in this case, the old slow-roll approximation performs better than the new ones. This scenario is fundamentally different from what was presented in Ref.~\cite{We}.

    The left panel shows the evolution of the slow-roll parameter \(\varepsilon_1\) near the end of inflation as a function of the inflaton field \(\phi\). The black curve corresponds to the numerical solution of the full dynamical system~(\ref{eq:numerical-1}), the blue line represents the standard slow-roll approximation, the red line corresponds to Approximation~I, and the green line to Approximation~II. The end of inflation is defined by the condition \(\varepsilon_1 = 1\), which coincides with the upper boundary of the plot.  

    It is evident from the figure that throughout the inflationary phase, the blue curve (standard slow-roll approximation) remains significantly closer to the black curve (numerical solution) than the curves associated with the other approximations.

    We now turn to the right panel of the figure, which displays the value of the slow-roll parameter \(\varepsilon_1\) at the onset of inflation. As can be seen, the same trend is already evident for large values of $\phi$: the standard (classical) slow-roll approximation describes the behavior of the system considerably better than the new approximations.  

    The plot also marks the points corresponding to $65$ and $60$ e-foldings, respectively. Since different approximations predict the end of inflation for different values of $\phi$, a fixed number of e-foldings corresponds to different initial values of the scalar field. We see that again the standard approximation consistently outperforms the newer ones.

\begin{figure}
\includegraphics[scale=0.44]{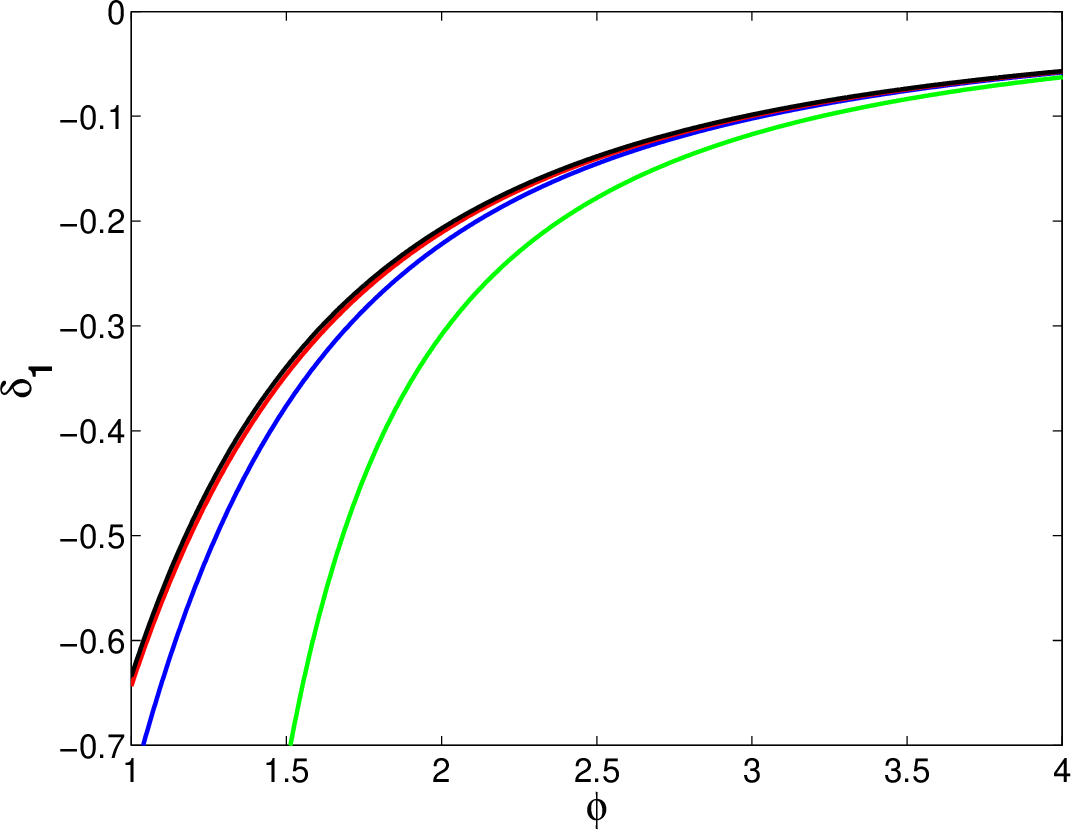}~~~~
\includegraphics[scale=0.44]{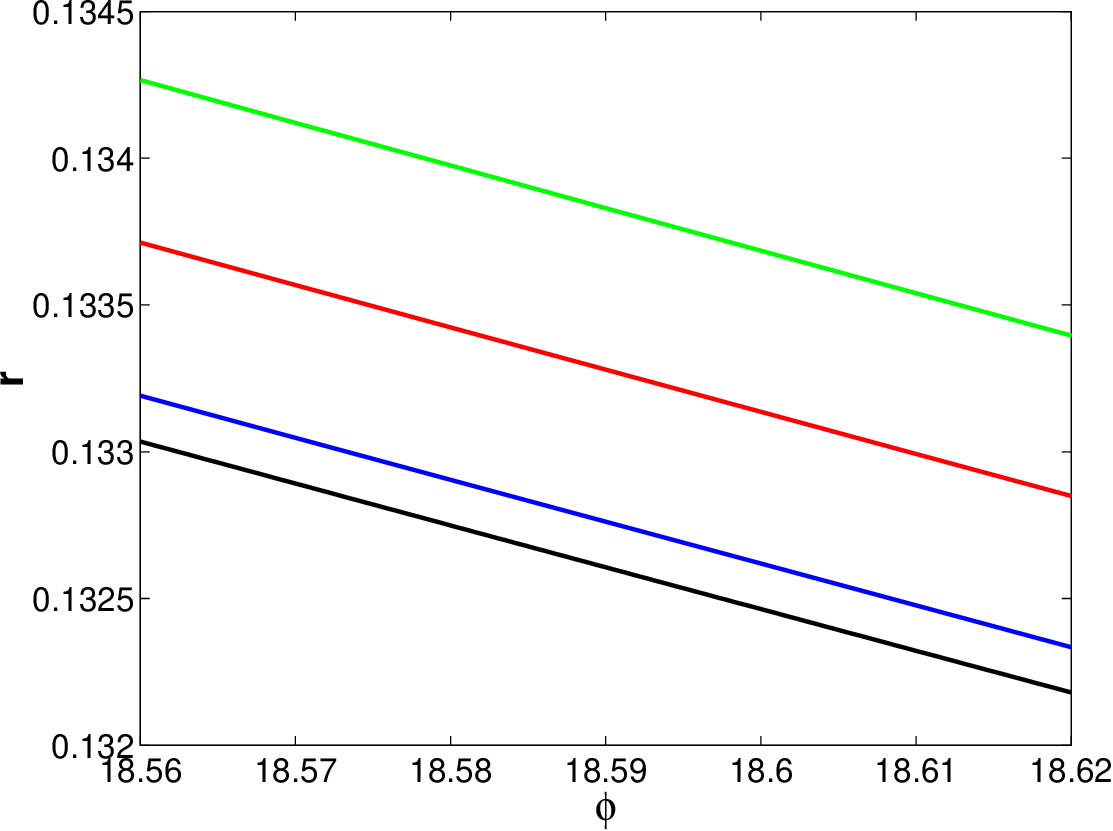}
\caption{\textit{On left panel}: The evolution of the slow-roll parameter $\delta_1(\phi)$ at the end of inflation for the Example~1. \textit{On right panel}: The plot of the tensor-to-scalar ratio \( r \) at the start of inflation for the Example~1.}
\label{Fig3}
\end{figure}

    The situation is somewhat different for the slow-roll parameter \(\delta_1\). To illustrate this, we refer to the left panel of Fig.~\ref{Fig3} which shows the behavior of \(\delta_1\) near the end of inflation. Here, first new approximation describes the parameter more accurately than both the standard slow-roll approximation and the second new approximation, as expected. However, the second new approximation still performs worse than the standard one.

\begin{figure}
\includegraphics[scale=0.44]{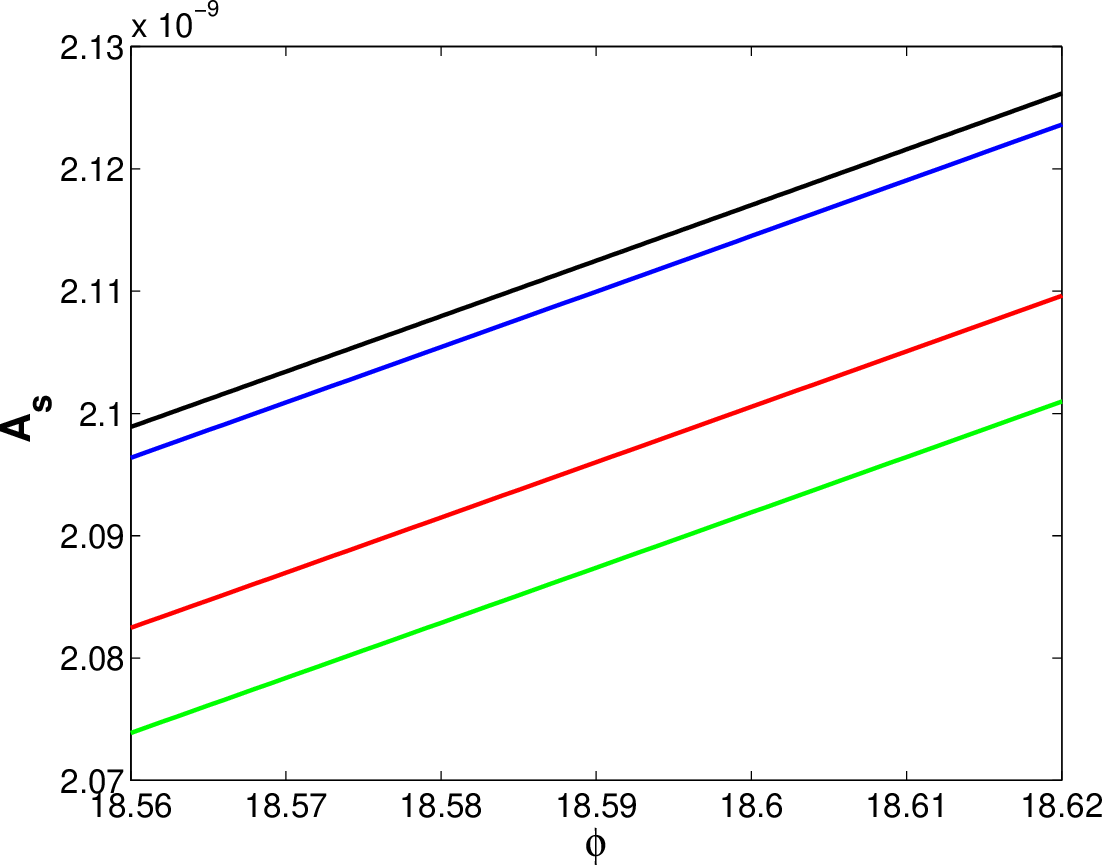}~~~~
\includegraphics[scale=0.44]{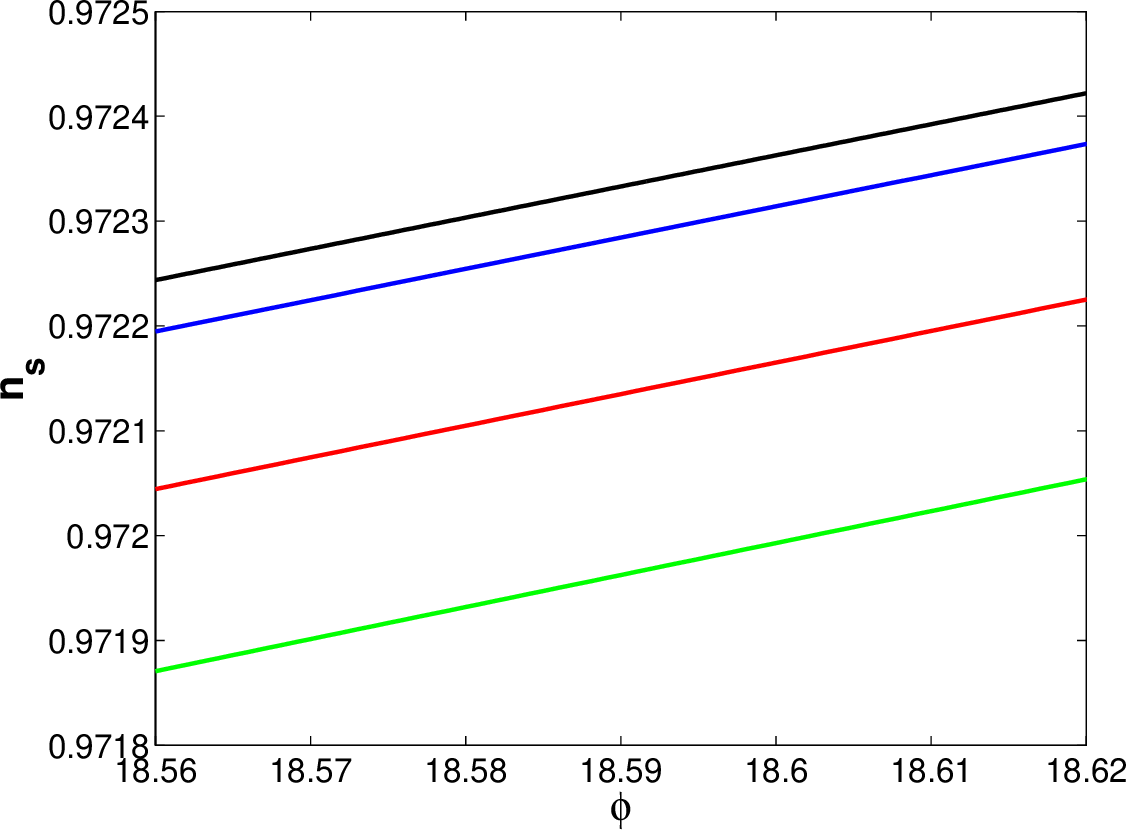}
\caption{The plot of the inflation parameter $A_s(\phi)$~(left) and $n_s(\phi)$~(right) at the beginning of inflation for the Example~1. As the plots show, the old approximation performs better than the new ones.}
\label{Fig4}
\end{figure}

        For the observable inflationary parameters --- the tensor-to-scalar ratio \(r\) (right panel of Fig.~\ref{Fig2}), the scalar amplitude \(A_s\) and the scalar spectral index \(n_s\) in Fig.~\ref{Fig4} --- the standard slow-roll approximation yields better agreement with the numerical solution than either of the two new approximations.

\begin{figure}
\includegraphics[scale=0.46]{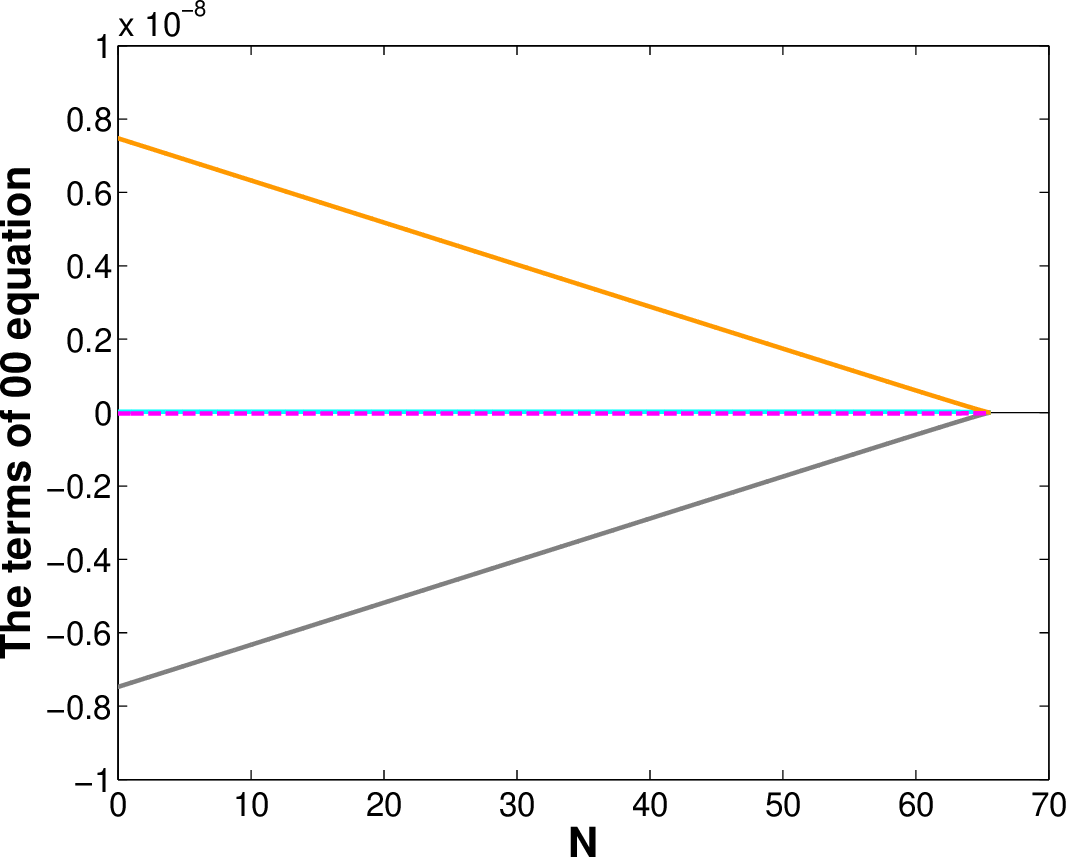}~~~~
\includegraphics[scale=0.46]{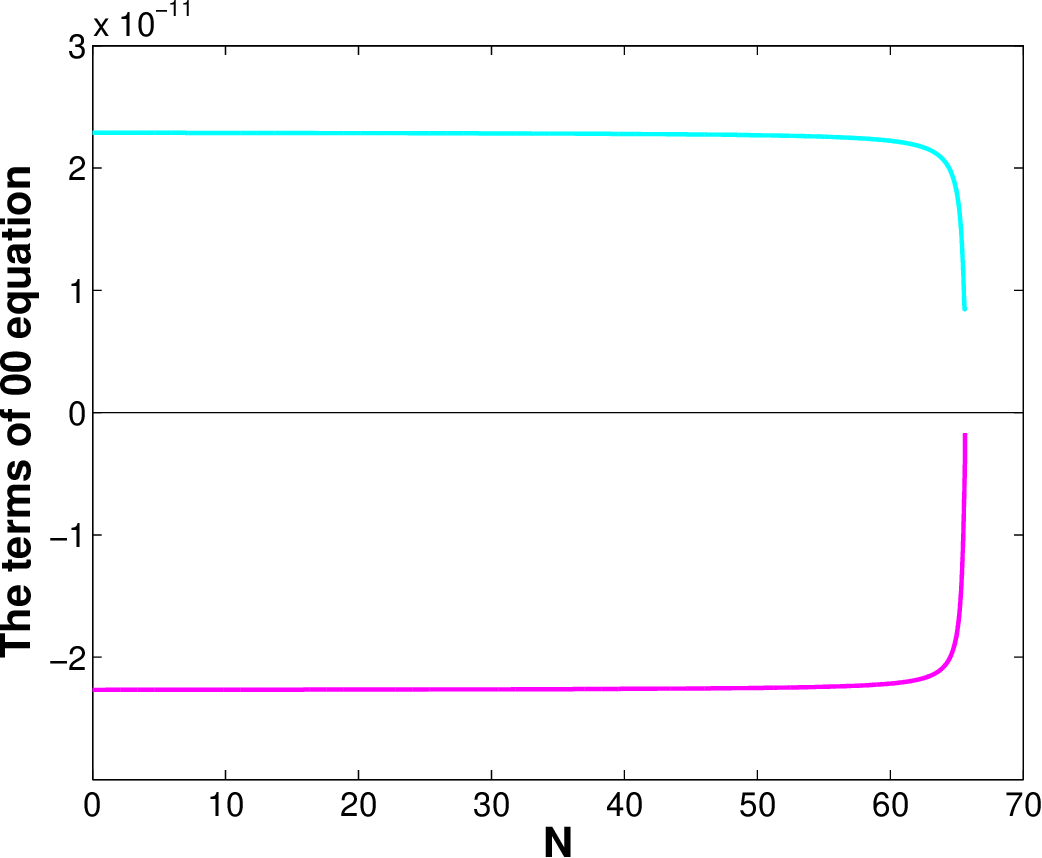}
    \caption{The numerical evolution of the different terms of dynamical equations as a functions of the number of e-foldings $N$ for the Example~1, where $-12UH^2$ is plotted in gray, ${\dot\phi}^2$ --- cyan, $2V$ --- orange, $24\xi_{,\phi}\dot\phi H^3$ --- magenta. In the left plot we see main terms, in the right plot --- corrections.}
    \label{Fig5}
\end{figure}

    To understand the origins of this behavior, we examine the behavior of different terms in the full system of equation of motion numerically. For bigger scale (\(10^{-8}\)), in the left panel of Fig.~\ref{Fig5}, two dominant terms are presented: the potential term \(2V\) and the term \(-12\,U H^2\). The smaller scale showing corrections (\(10^{-10}\)) presented at the right panel of Fig.~\ref{Fig5} shows the behavior of the terms \(- 24\xi_{,\phi} \dot{\phi}H^3\) and \( \dot{\phi}^2\). The first term does enter into new approximations while the second does not. As the plot demonstrates, throughout the inflationary phase, these terms for the model in question have, however, nearly equal magnitudes but opposite signs. As in the full equation (\ref{eq4}) these contributions enter with the same sign, this leads to a near-cancellation in the final expression for \(\varepsilon_1\). Consequently, the attempt to improve the approximation by including additional terms actually makes it worse.

    Since all other inflationary observable parameters (\(r\), \(A_s\), \(n_s\), etc.) are analytically expressed in terms of \(\varepsilon_1\) (and related slow-roll parameters), this cancellation effect explains the observed feature in the figures: the standard slow-roll approximation --- which retains fewer terms turns out to be more accurate than the higher-order approximations for this specific model.

\subsubsection{Example 2}
~~~~We consider the same coupling function (\ref{xi1}) with
\begin{equation}
\label{ex2paramxi}
\phi_0 = 10^{-6},~~~~\alpha=\sqrt{10}\cdot 10^{-6},
\end{equation}
but now with a flatter potential \begin{equation}
\label{V2}
V(\phi)=V_0\tanh^2\left(\frac{\phi}{\phi_1}\right),~~~V_0 = 1.6\cdot 10^{-11},~~~~\phi_1=1.
\end{equation}
    In this example the potential is chosen to be asymptotically flat, this should turn the term \( \dot{\phi}^2\) to be less important. The plot of the terms Fig.~\ref{Fig8} shows that indeed these two terms, although being of the opposite signs, differ in magnitude. However, after calculating observable parameters we can see from the plot Fig.~\ref{Fig7} that the new approximations perform worse than the standard approximation in this case too.
 
\begin{figure}
\includegraphics[scale=0.46]{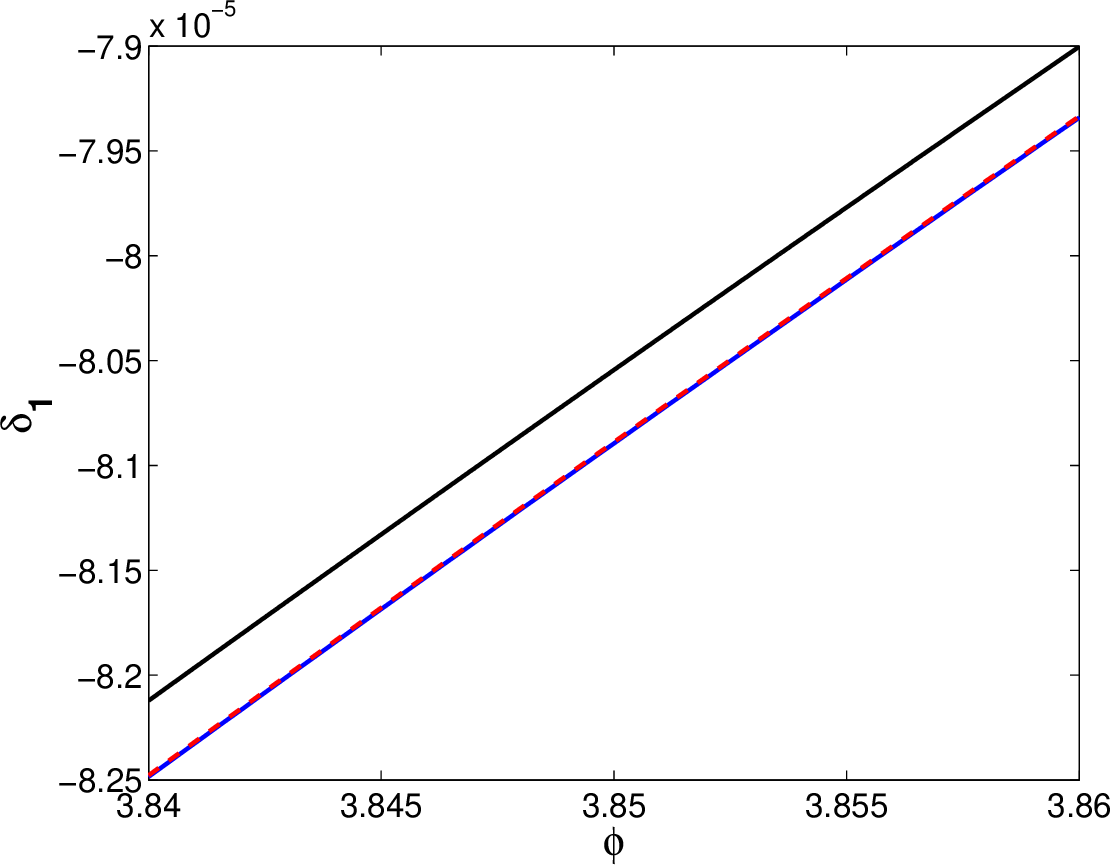}
\includegraphics[scale=0.46]{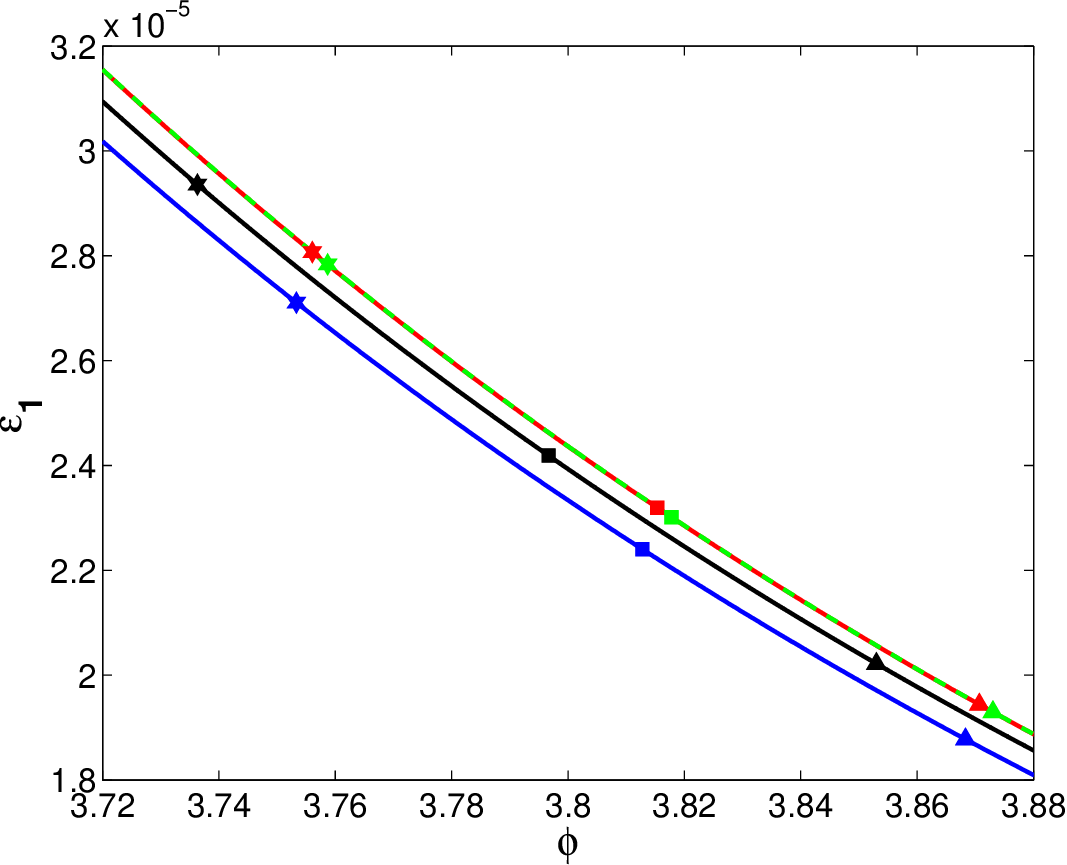}
    \caption{\textit{On left panel}: The evolution of the slow-roll parameter $\delta_1(\phi)$ at the beginning of inflation near $65$ e-foldings for the Example~2. All three approximations are practically indistinguishable. \textit{On right panel}: The evolution of the slow-roll parameters $\varepsilon_1$ at the start of inflation for the Example~2. For comparative purposes, the plot indicates different numbers of e-foldings: \(N = 65\) (triangles), \(N = 60\) (squares), and \(N = 55\) (stars). For visual clarity, the green line representing the II~new slow-roll approximation is shown as a dashed line.}
    \label{Fig6}
\end{figure}

\begin{figure}
\includegraphics[scale=0.3]{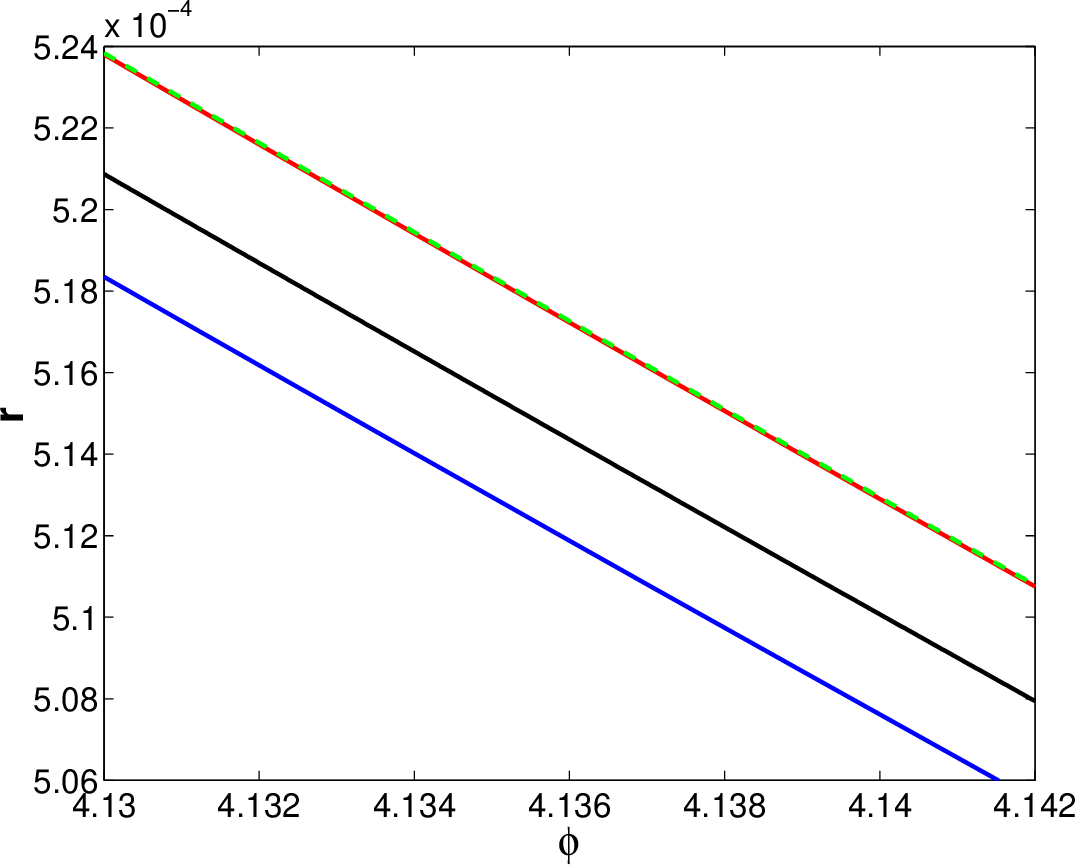}~
\includegraphics[scale=0.3]{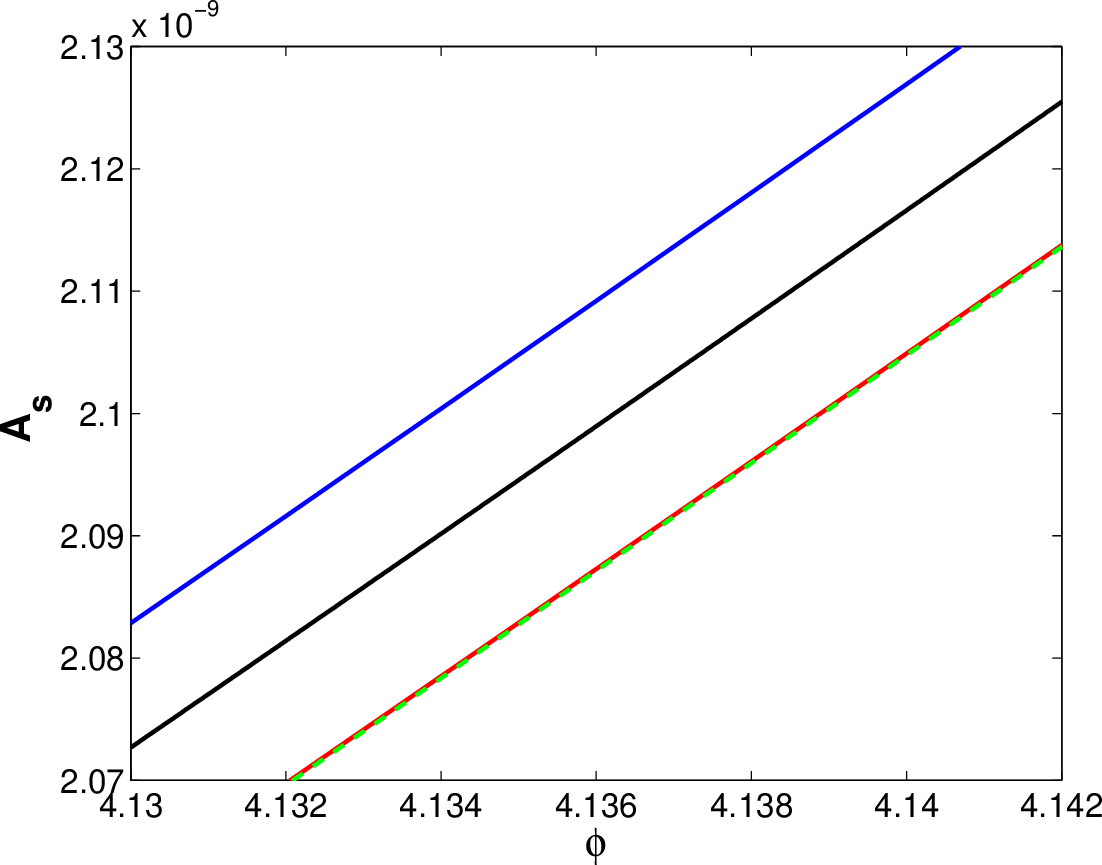}~
\includegraphics[scale=0.3]{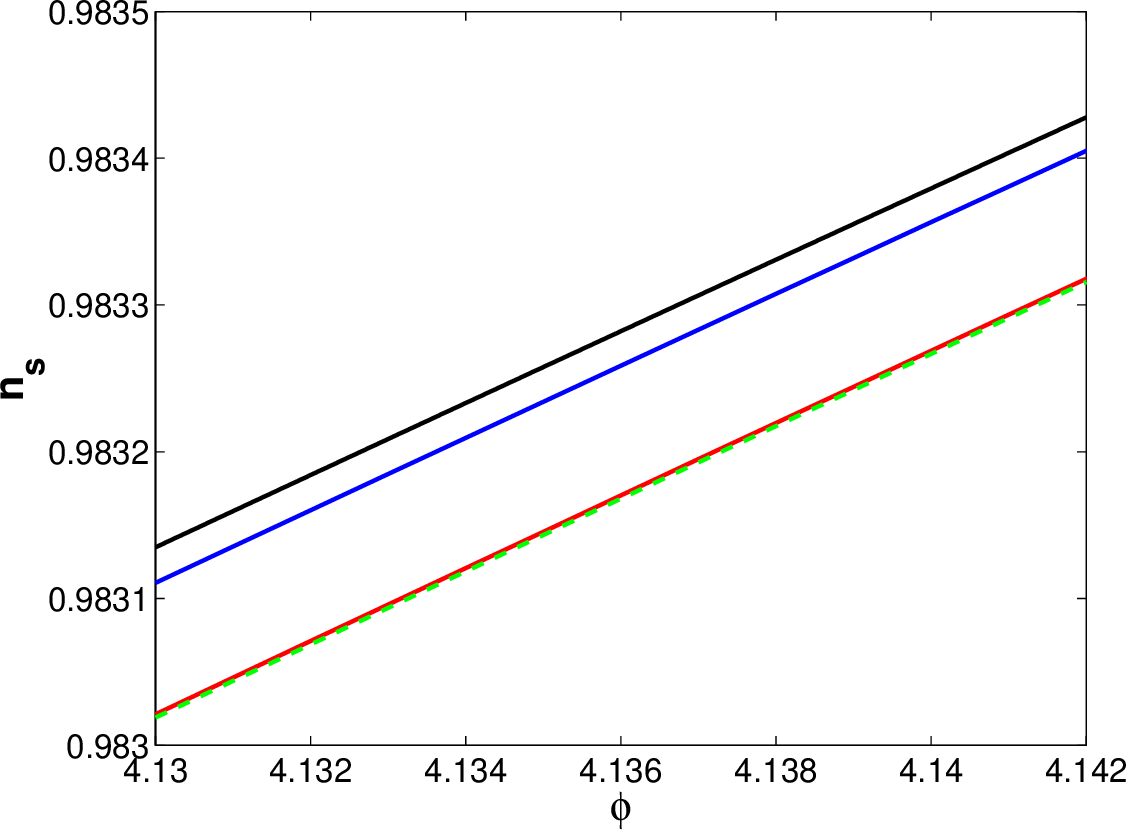}
\caption{The figure presents the evolution of the inflationary parameters $r$~(left), $A_s$~(middle), $n_s$~(right) at the beginning of inflation for the Example~2. As the plots show, the old approximation for \( n_s \) performs clearly better than the new ones.}
\label{Fig7}
\end{figure}

\begin{figure}
\includegraphics[scale=0.46]{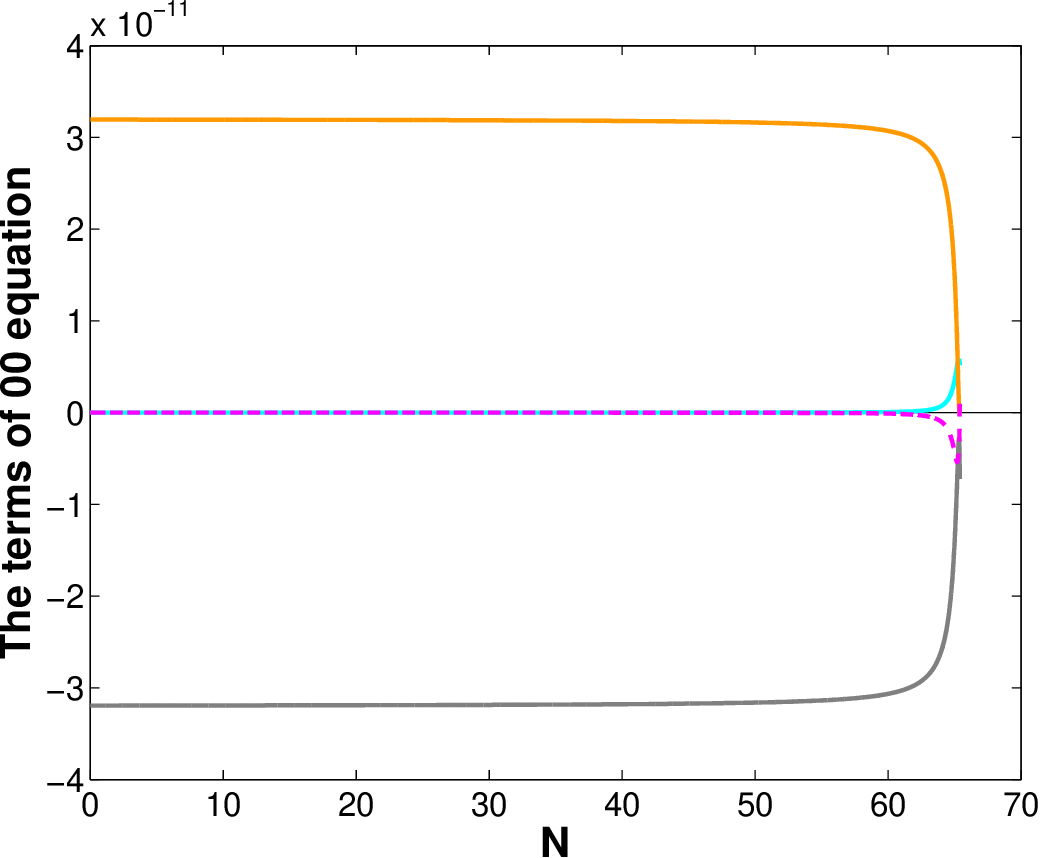}~~~~
\includegraphics[scale=0.46]{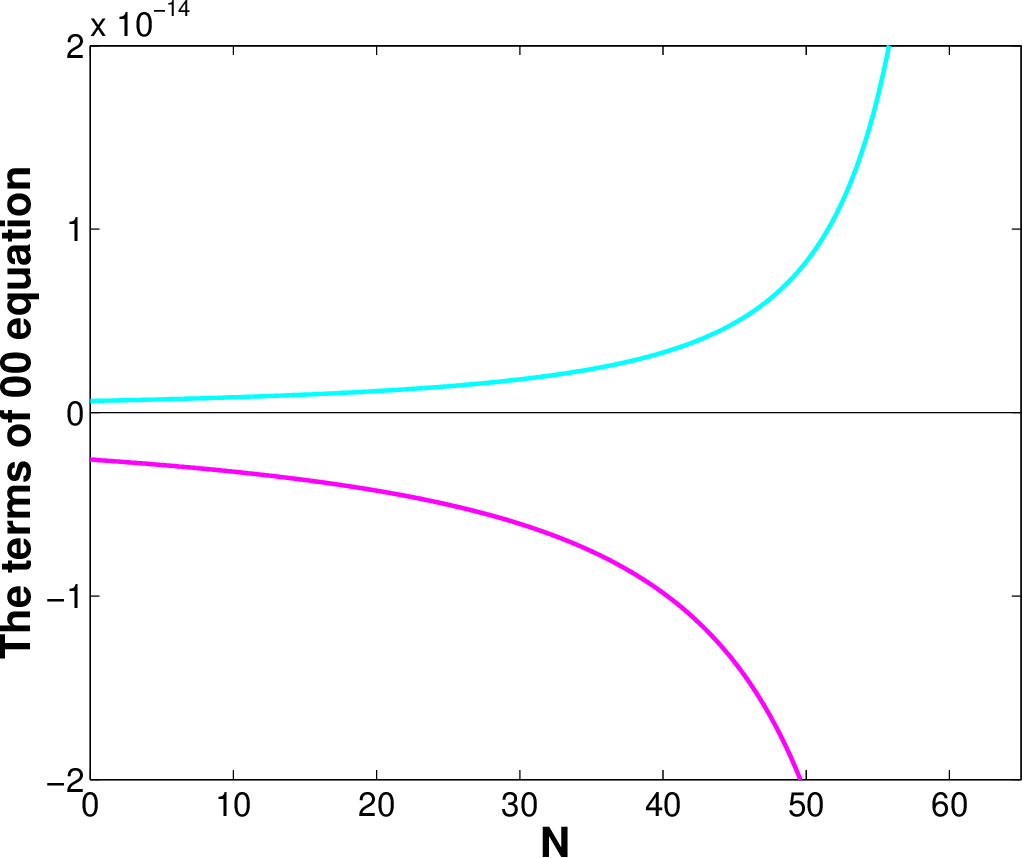}
    \caption{The numerical evolution of different terms of dynamical equations as a function of the number of e-foldings $N$ for the Example~2, where $-12UH^2$ is plotted in gray, ${\dot\phi}^2$ --- cyan, $2V$ --- orange, $24\xi_{,\phi}\dot\phi H^3$ --- magenta. In the left plot we see main terms, in the right plot --- corrections.}
    \label{Fig8}
\end{figure}

    The mechanism behind this phenomenon should be thus entirely different from the previous one. To understand the situation we plot the slow-roll parameters \(\varepsilon_1\) and \(\delta_1\) at the beginning of inflation (Fig.~\ref{Fig6}). For \(\varepsilon_1\), at the start of inflation, the old approximations perform worse, as we can anticipate, and it can be seen clearly from~Fig.~\ref{Fig6}. Note the sign of the difference between the approximations and the numerical solution. For the old approximation it is negative, while for the new ones it is positive. As for $\delta_1$ all approximations have roughly the same effectiveness, and the difference between all three approximation and the numerical solution is negative. This explains the lower effectiveness of the new approximations for calculating of $n_s$: in the formula for it~(\ref{eq:r}), the slow-roll parameters enter with different signs, so if the signs of errors in $\varepsilon_1$ and $\delta_1$ are the same (as in the old approximation), the errors subtract and to some extend compensate each other, while if they are opposed (as in the new approximations), the errors sum up.

    It should be noted that the set of constants $V_0$, $\phi_0$, $\alpha$ for this particular example is specially chosen to get this seemingly strange result --- better fitting of spectral index reproduces the general situation when old and new approximations give similar precision.

\subsection{When new approximations are needed}
\begin{figure}
\includegraphics[scale=0.3]{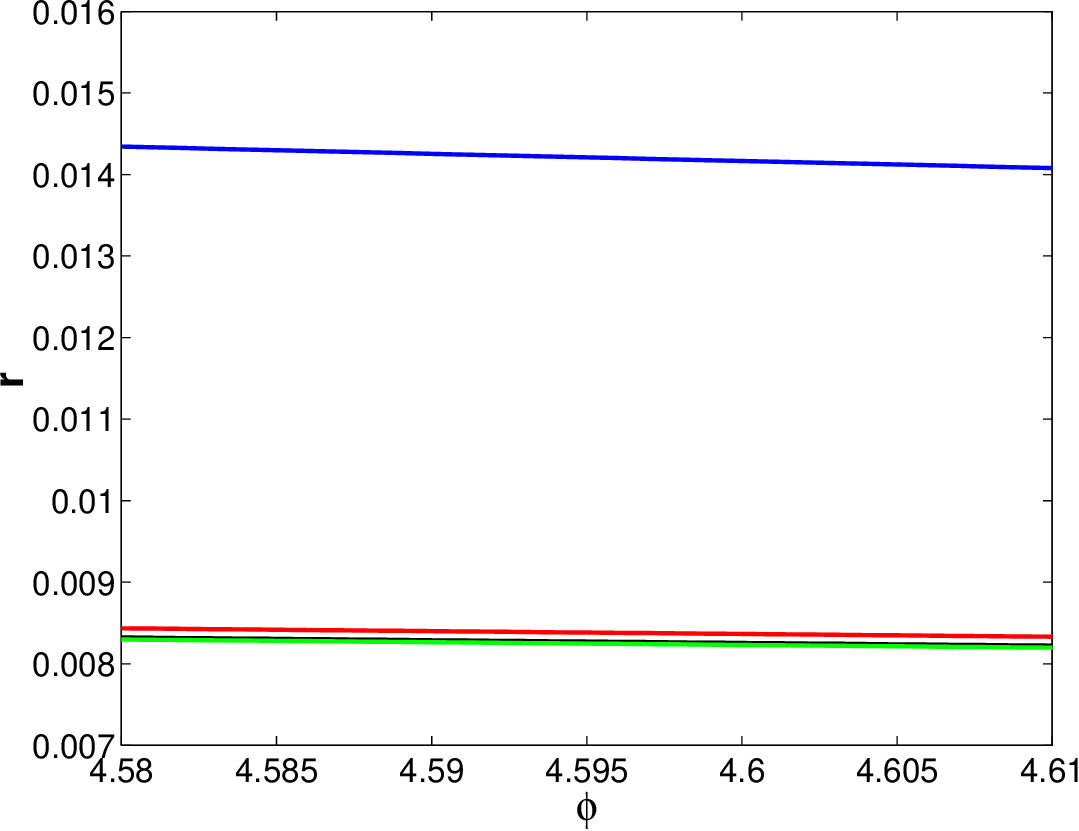}~
\includegraphics[scale=0.3]{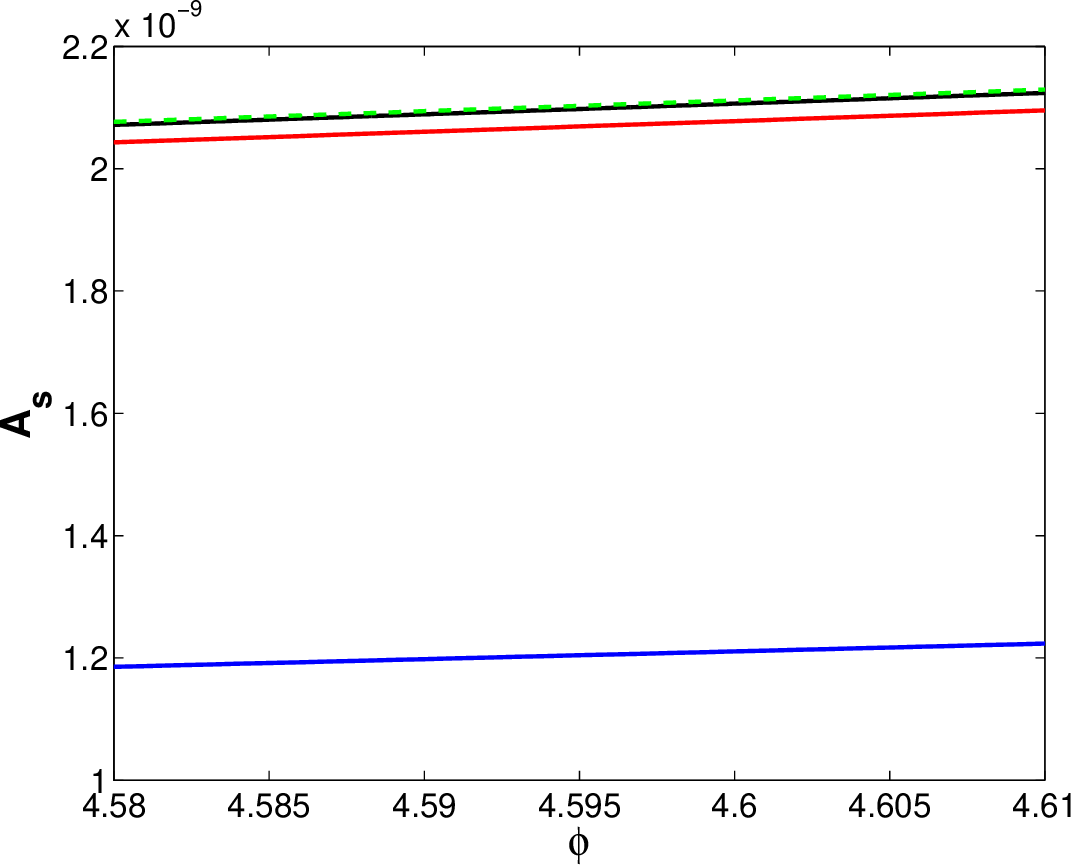}~
\includegraphics[scale=0.3]{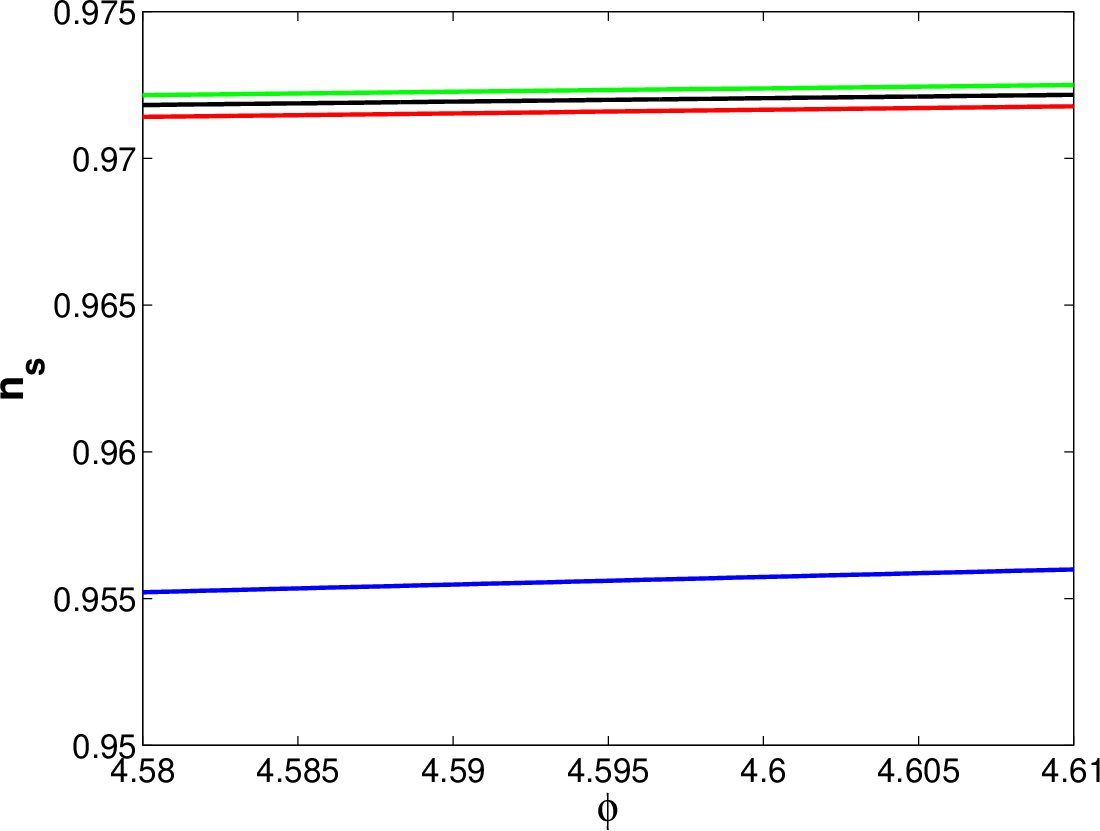}
\caption{The figure presents the evolution of the inflationary parameters $r$~(left), $A_s$~(middle), $n_s$~(right) at the beginning of inflation near $69$ e-foldings for the model (\ref{xi-}), (\ref{paramxi-}), (\ref{V}), (\ref{paramV3}). This is example of the realistic scenario.}
\label{Fig9}
\end{figure}

~~~~Though these cases can be considered as rather exotic, they give us a possible clue to the problem of better slow-roll approximation in Gauss-Bonnet cosmology. Namely, signs of the terms $\phi^2$ and $24\xi_{,\phi} \dot{\phi}H^3$ in both cases are different. In order to check the importance of these signs we finally consider a regular coupling function but with negative sign:
\begin{equation}
\label{xi-}
\xi(\phi)=-\left(\frac{\phi^2}{{(\alpha\phi)}^2+\phi_0^2 }\right),
\end{equation}
where it is chosen
\begin{equation}
\label{paramxi-}
\alpha=2.04 \cdot 10^{-6},~~~~\phi_0=10^{-6},
\end{equation}
and a quadratic potential (\ref{V}) with
\begin{equation}
\label{paramV3}
V_0=1.2\cdot 10^{-11}.
\end{equation}
 In this case $24\xi_{,\phi}\dot{\phi}H^3$ changes its sign while $\dot{\phi}^2$ term evidently keeps it. One example is presented in Fig.~\ref{Fig9}. Note that despite the potential is a quadratic one, the influence of the Gauss-Bonnet term allows to match both observational parameters as well as current bounds on $r$. We see that for this model new approximations work considerably better than the standard one. So that, we think that detected earlier poor performance of the standard approximation is connected not with a particular rather specific form of the coupling function $\xi(\phi)$ chosen in \cite{We} but with the sign of $\xi_{,\phi}$. When it is positive, keeping the term $24\xi_{,\phi} \dot{\phi}H^3$ and neglecting $\dot{\phi}^2$ is not reasonable since they are of different signs, and new approximations cannot significantly influence accuracy. On the other hand, in the case of a negative $\xi_{,\phi}$ it is better to use more involved approximations.

\section{Conclusion}
~~~~In the present paper we have considered three slow-roll approximations described earlier~\cite{We} for Gauss-Bonnet inflation when behavior of the coupling function at the Gauss-Bonnet term is different from considered in~\cite{We}. In the cited paper the case of coupling function being roughly inversely proportional to the scalar field potential have been studied, and it appears that the standard slow-roll approximation~\cite{VanDeBruck:2016jll} reproduces actual dynamics rather badly, so more involved approximations are needed.

    Here we start with a situation with both potential and coupling functions being increasing functions of the scalar field, and the resulting the Gauss-Bonnet corrections are small with respect to the main terms. A typical situation appears to be as presented by Fig.~\ref{Fig1} of the present paper. New approximations give almost no additional precision with respect the the older one. This means that in such cases there is no need to use more involved formulas.

    What appears to be rather surprising is that in some cases the old approximation is not only as good as new ones, but is actually better. This needs explanation, and in our paper we present the corresponding analysis for two models. One of which (the model~1) represent a situation when omitted term proportional to ${\dot\phi}^2$ is of the same order as the term included in the new approximation and having different sign, almost canceling it. Another model (the model~2) shows even stranger situation when despite slow-roll parameters are better approximated by new approximations, the observation oriented quantities $n_s$ and $r$ are better approximated by the old approximation.

After that we consider the coupling function which is a decreasing function of the scalar field, but, in contrast to \cite{We} we still use a positive power index but change the sign of the function. After the sign flips, the standard approximation appears to be significantly less accurate than the new ones. Summarizing, our results might indicate that new more cumbersome approximations should be used if $\xi_{,\phi}<0$, while for $\xi_{,\phi}>0$ the standard approximations works as well as (and in some particular cases even better than) the new ones.

\subsection*{Acknowledgements}
~~~~Authors are grateful to Ekaterina Pozdeeva and Sergey Vernov for discussions. The study was conducted under the state assignment of Lomonosov Moscow State University.


\begin{thebibliography}{99}
\bibitem{We}
E.~O.~Pozdeeva, M.~A.~Skugoreva, A.~V.~Toporensky and S.~Yu.~Vernov,
JCAP {\bf 09}, 050 (2024).

\bibitem{Liddle1994}
A.~R.~Liddle, P.~Parsons and J.~D.~Barrow,
Phys.\ Rev.\ D {\bf 50}, 7222 (1994).

\bibitem{Akin:2020qjw}
K.~Akin, S.~Arapo\v{g}lou and A.~E.~Y\"{u}kselci,
Phys.\ Dark\ Univ. {\bf 30}, 100691 (2020).

\bibitem{Starobinsky1979}
A.~A.~Starobinsky,
JTEP\ Lett. {\bf 30}, 682 (1979).

\bibitem{Guth1981}
A.~H.~Guth,
Phys.\ Rev.\ D {\bf 23}, 347 (1981).

\bibitem{Linde1982}
A.~D.~Linde,
Phys.\ Lett.\ B {\bf 108}, 389 (1982).

\bibitem{Linde1983}
A.~D.~Linde,
Phys.\ Lett.\ B {\bf 129}, 177 (1983).

\bibitem{Mukhanov1981}
V.~F.~Mukhanov and G.~V.~Chibisov,
JTEP\ Lett. {\bf 33}, 532 (1981).

\bibitem{PlanckCollaboration2023}
R.~L.~Workman {\it et al.} (Planck Collaboration), Review of Particle Physics, PTEP {\bf 2022} (2022) 083C01.

\bibitem{Durrer2015}
R.~Durrer,
C.\ R.\ Physique {\bf 16} 948 (2015).

\bibitem{Abazajian2015}
K.~N.~Abazajian {\it et al.},
Astroparticle Physics {\bf 63} 55 (2015).

\bibitem{Louis2025}
T.~Louis {\it et al.} (ACT Collaboration),
[arXiv:2503.14452].

\bibitem{Calabrese2025}
E.~Calabrse {\it et al.} (ACT Collaboration),  
[arXiv:2503.14454].

\bibitem{DESICollaboration2024} 
A.~G.~Adame {\it et al.} (DESI Collaboration),
[arXiv:2404.03002].

\bibitem{Capozziello2011}
S.~Capozziello and M.~De~Laurentis,
Phys.\ Rept. {\bf 509} 167 (2011).

\bibitem{Lombriser2011}
L.~Lombriser,
{\it Confronting Theories of Gravity with Large-Scale Structures}, Ph.D. thesis,
Universit\"{a}t Z\"{u}rich, 2011.

\bibitem{Akrami2019}
Y.~Akrami {\it et al.} (Planck Collaboration),
Astron.\ Astrophys. {\bf 641} A10 (2020).

\bibitem{BICEP:2021xfz}
P.~A.~R.~Ade {\it et al.} (BICEP/Keck Collaboration),
Phys.\ Rev.\ Lett. {\bf 127} 151301 (2021).

\bibitem{Galloni:2022qjw}
G.~Galloni, N.~Bartolo, S.~Matarrese, M.~Migliaccio, A.~Ricciardone and N.~Vittorio,
JCAP {\bf 04}, 062 (2023).

\bibitem{Guo:2010jr}
Z.-K.~Guo and D.~J.~Schwarz,
Phys.\ Rev.\ D {\bf 81}, 123520 (2010).

\bibitem{Jrv2022}
L.~J\"{a}rv and A.~Toporensky
Eur.\ Phys.\ J.\ C {\bf 82}, 179 (2022).

\bibitem{Chern1944}
S.-S.~Chern,
Annals of Mathematics {\bf 45}, 747 (1944).

\bibitem{Jiang2013}
P.-X.~Jiang, J.-W.~Hu and Z.-K.~Guo,
Phys.\ Rev.\ D {\bf 88}, 123508 (2013).

\bibitem{VanDeBruck:2016jll}
C.~van~de~Bruck and C.~Longden,
Phys.\ Rev.\ D {\bf 93}, 063519 (2016).

\bibitem{Pozdeeva2021}
E.~O.~Pozdeeva and S.~Yu.~Vernov,
Eur.\ Phys.\ J.\ C {\bf 81}, 633 (2021).

\bibitem{Pozdeeva:2019akw}
E.~O.~Pozdeeva, M.~Sami, A.~V.~Toporensky and S.~Yu.~Vernov,
Phys.\ Rev.\ D {\bf 100}, 083527 (2019).

\bibitem{Koh:2016abw}
S.~Koh, B.-H.~Lee and G.~Tumurtushaa,
Phys.\ Rev.\ D {\bf 95}, 123509 (2017).

\bibitem{Odintsov:2023qjw}
S.~D.~Odintsov and T.~Paul,
Phys.\ Dark\ Univ. {\bf 42}, 101263 (2023).

\bibitem{Gasperini_1997}
M.~Gasperini, M.~Maggiore and G.~Veneziano,
Nucl.\ Phys.\ B {\bf 494}, 315 (1997).

\bibitem{Neupane_2006}
I.~P.~Neupane and B.~M.~N.~Carter
JCAP {\bf 06}, 004 (2006).

\bibitem{Tsujikawa_2002}
S.~Tsujikawa, R.~Brandenberger and F.~Finelli,
Phys.\ Rev.\ D {\bf 66}, 083513 (2002).

\bibitem{Antoniadis:1993jc}
I.~Antoniadis, J.~Rizos and K.~Tamvakis,
Nucl.\ Phys.\ B {\bf 415}, 497 (1994).

\bibitem{Cartier:2001is}
C.~Cartier, J.-c.~Hwang and E.~J.~Copeland,
Phys.\ Rev.\ D {\bf 64}, 103504 (2001).

\bibitem{Kawai:1998ab}
S.~Kawai, M.-a.~Sakagami and J.~Soda,
Phys.\ Lett.\ B {\bf 437}, 284 (1998).

\bibitem{Kawai:1999pw}
S.~Kawai and J.~Soda,
Phys.\ Lett.\ B {\bf 460}, 41 (1999).

\bibitem{Torii:1996xd}
T.~Torii, H.~Yajima and K.-i.~Maeda,
Phys.\ Rev.\ D {\bf 55}, 739 (1997).

\bibitem{Hwang:2004th}
J.-c.~Hwang and H.~Noh,
Phys.\ Rev.\ D {\bf 71}, 063536 (2005).

\bibitem{Sami:2005zc}
M.~Sami, A.~Toporensky, P.~V.~Tretjakov and S.~Tsujikawa,
Phys.\ Lett.\ B {\bf 619}, 193 (2005).

\bibitem{Tsujikawa:2006tv}
S.~Tsujikawa and M.~Sami,
JCAP {\bf 01}, 006 (2007).

\bibitem{Bahamonde:2017ize}
S.~Bahamonde, C.~G.~B\"{o}hmer, S.~Carloni, E.~J.~Copeland, W.~Fang and N.~Tamanini,
Phys.\ Rep. {\bf 775}, 1 (2018).

\bibitem{Zhu2018}
Q.~Wu, T.~Zhu and A.~Wang,
Phys.\ Rev.\ D \textbf{97}, 103502 (2018).

\bibitem{Zhu2023}
T.~C.~Li, T.~Zhu and A.~Wang,
JCAP \textbf{05}, 006 (2023).

\bibitem{Zhu2025}
Yogesh, A.~Mohammadi, Q.~Wu and T.~Zhu,
JCAP \textbf{10}, 010 (2025).

\bibitem{Odintsov2020}
S.~D.~Odintsov, V.~K.~Oikonomou and F.~P.~Fronimos,
Nucl.\ Phys.\ B \textbf{958}, 115135 (2020).

\bibitem{Odintsov2025} 
S.~D.~Odintsov and V.~K.~Oikonomou, 
Phys.\ Lett.\ B \textbf{868}, 139779 (2025).

\bibitem{Oikonomou2021a}
V.~K.~Oikonomou and F.~P.~Fronimos,
Class.\ Quant.\ Grav. \textbf{38}, 035013 (2021).

\bibitem{Oikonomou2021b}
V.~K.~Oikonomou,
Class.\ Quant.\ Grav. \textbf{38}, 195025 (2021).

\bibitem{Oikonomou2022a}
V.~K.~Oikonomou,
Astropart.\ Phys. \textbf{141}, 102718 (2022).

\bibitem{Oikonomou2022b}
V.~K.~Oikonomou, P.~D.~Katzanis and I.~C.~Papadimitriou,
Class.\ Quant.\ Grav. \textbf{39}, 095008 (2022).

\bibitem{Oikonomou2024}
V.~K.~Oikonomou,
Phys.\ Lett.\ B \textbf{856}, 138890 (2024).

\bibitem{Oikonomou2025}
V.~K.~Oikonomou,
[arXiv:2508.19196].

\bibitem{Yogesh2025a}
Yogesh and Abolhassan~Mohammadi,
Astrophys.\ J. {\bf 986}, no.1, 35 (2025).

\bibitem{Yogesh2025b}
Yogesh,~Mehnaz~Zahoor, Kashif~Ali~Wani and Imtiyaz~Ahmad~Bhat,
Phys.\ Dark Univ. {\bf 47}, 101732 (2025).

\bibitem{Yogesh2025c}
Yogesh, Imtiyaz~Ahmad~Bhat and Mayukh~R.~Gangopadhyay,
Phyd.\ Dark Univ. {\bf 49}, 102021 (2025).

\bibitem{We1}
E.~O.~Pozdeeva, M.~A.~Skugoreva, A.~V.~Toporensky and S.~Yu.~Vernov,
JCAP {\bf 05}, 081 (2025).

\bibitem{Pozdeeva:2020zpu}
E.~O.~Pozdeeva, M.~R.~Gangopadhyay, M.~Sami, A.~V.~Toporensky and S.~Yu.~Vernov,
Phys.\ Rev.\ D {\bf 102}, 043525 (2020).

\bibitem{Sudarsky}  
R.-L.~Lechuga and  D.~Sudarsky, 
[arXiv:2502.05393].
\end{thebibliography}
\end{document}